\documentclass[aps,prl,twocolumn,groupedaddress,superscriptaddress,showpacs]{revtex4-1}

\usepackage{amssymb}
\usepackage{amsmath}
\usepackage{latexsym}
\usepackage{revsymb}

\newcommand{\ZZ}{\mathbb{Z}}
\newcommand{\Sph}{\mathbb{S}}
\newcommand{\NN}{\mathbb{N}}
\newcommand{\MM}{\mathbb{M}}

\begin{document}

\title{
Topological phase states of the $SU(3)$ QCD
}

\author{Alexander P. Protogenov}
 \affiliation{Institute of Applied Physics, Nizhny Novgorod 603950,
Russia}
 \affiliation{ Donostia International Physics Center
(DIPC), 20018 San Sebasti\'an/Donostia, Spain}

\author{Evgueni V. Chulkov}
\affiliation{Donostia International Physics Center (DIPC),
20018, San Sebasti\'an/Donostia, Spain}
\affiliation{Departamento de F\'{\i}sica de Materiales UPV/EHU,
Centro de F\'{\i}sica de Materiales CFM - MPC and Centro Mixto
CSIC-UPV/EHU, 20080 San Sebasti\'an/Donostia, Spain}

\author{Jeffrey C. Y. Teo}
 \affiliation{Department of Physics, University of Illinois at Urbana-Champaign, Urbana IL 61801, USA}

\begin{abstract}
We consider the topologically nontrivial phase states and the corresponding topological defects in the $SU(3)$ $d$-dimensional quantum chromodynamics (QCD). The homotopy groups for topological classes of such defects are calculated explicitly. We have shown that the three nontrivial groups are $\pi_{3}SU(3)=\ZZ$, $\pi_{5}SU(3)=\ZZ$, and $\pi_{6}SU(3)=\ZZ_{6}$ if $3 \leq d \leq 6$. The latter result means that we are dealing exactly with six topologically different phase states. The topological invariants for d=3,5,6 are described in detail.
\end{abstract}

\maketitle

 {\it Introduction}. -  Topological invariants of
field configurations are the fundamental objects in the quantum field theory and condensed matter physics, which  classify topological defects and possible phase states \cite{Mer, Vol}.
The well-known examples of topological field distributions are vortices, hedgehogs, and instantons. They are a direct consequence of the nontrivial homotopy groups $\pi_{n}{\Sph}^{n}={\ZZ}$ for the spheres ${\Sph}^{n}$ with $n=1,2$, $\pi_{3}SU(2)=\ZZ$,
 and $d=1,2,3$, respectively. Here, $\ZZ$ is a group of integers. Recent progress in the theory of topologically ordered phase states \cite{TeoKane,TeoKane2} is associated with the classification of the systems, in which the D-dimensional surface ${\Sph}^{D}$ surrounds a defect in d-dimensional topological insulators or superconductors \cite{HasanKane,QiZhang,Schnyder,Kitaev}  and $D \neq d$. In this case, the first nontrivial example $\pi_{3}{\Sph}^{2}=\ZZ$ is the well-known Hopf mapping of the three-dimensional sphere ${\Sph}^{3}$ into the two-dimensional one ${\Sph}^{2}$. The corresponding topological Hopf invariant has the form $Q=\frac{1}{16\pi^{2}}\int_{{\Sph}^{3}}A\wedge dA \in \ZZ$. It is called  the helicity in magnetohydrodynamics or the Abelian Chern-Simons action in the $(2+1)$-dimensional topological
 field theory. The integer $Q$ means the knotting degree of the field distributions and determines, in particular, the lower bound  \cite{VakKap,ProtVerb} of the energy of the two-component Ginzburg-Landau model expressed \cite{BabFad} in the form of the Skyrme-Faddeev-Niemi model \cite{Prot,JaykHietSalo}.
 In this $O(3)$ ${\sigma}$-model, the $U(1)$ two-form $dA={\bf n}[{\rm d}{\bf n}\wedge {\rm d}{\bf n}]$ is parametrized  by the unit 3d vector ${\bf n}$ which maps the base space ${\Sph}^{3}$ into ${\Sph}^{2}$. The target sphere ${\Sph}^{2}$ of the map is topologically equivalent to the coset
$SU(2)/U(1) \cong {\Sph}^{2}$. The ${\bf n}$-field is also a relevant on-shell variable \cite{Chi,Cho,FadNiemi,Fad} in the infrared limit of the $SU(2)$ QCD.

In this Letter, we use the $SU(3)$ group instead of the $SU(2)$ one.
The change in the value of the rank is due to several reasons.
Primarily, to the three colors of the QCD. From the point of view of the knot theory, this choice is also due to an attempt to extend the low-dimensional topology of the standard knot theory to higher dimensions of the $SU(3)$ QCD target space \cite{Fa}. One approach to this problem is to use many-valued functionals \cite{Novikov} in accordance with the conjecture given in Ref. \cite{Arnold}.
Another elegant method is based on the results obtained in Ref. \cite{Ran}.
However, it is more expedient to describe the target spaces of the $SU(3)$ QCD as a  generalization of the $SU(2)$ target sphere ${\Sph}^{2}$.

{\it Flag space}. - To generalize the $SU(2)$ group target space to the $SU(3)$ one,
we consider the coset  $SU(3)/(U(1)\times U(1))=F_{2}$ that is now
the flag space $F_{2}$  \cite{Pick,KondoTaira}. The remained freedom
of the maps is the dimension of the base space having a nontrivial homotopy group.
For simplicity, we restrict our consideration to the spheres ${\Sph}^{n}$ as the base spaces. Therefore, we focus on such $n$ of the homotopy groups $\pi_{n}F_{2}$, which
yields nontrivial results. For comparison, in addition to the maximal torus
$U(1)^{2}$ of $SU(3)$ that results in the general orbit $F_{2}$ \cite{BerHol}, we
calculate  the homotopy groups of the degenerate orbit $CP^{2}$, which are
equivalent to the coset space $SU(3)/U(2)=SU(3)/\left(SU(2)\times U(1)\right)=CP^{2}$.

It should be noted that we are restricted to the framework of the homotopy group approach. Therefore, we would like to determine the constraints on the type of the possible topological phase states and topological defects only. We will describe the geometry of the flag space $F_{2}$ and topological features in the last two sections. The results of our calculations are presented in two tables.

It is seen in Table I that the nontrivial homotopy groups for $d \leq 6$ are
$\pi_{2}F_{2}$, $\pi_{3}F_{2}$, $\pi_{5}F_{2}$, and $\pi_{6}F_{2}$.

\begin{table}
\centering
\begin{tabular}{c|cccccccccccc|ccccccccccc}
$d  $& \multicolumn{1}{c}{}
$0$ & $1$   &  $2$ &  $3$ &  $4$ &  $5$ & $6$ & $7$ & $8$ & $9$ & $10$ &           \\
\hline
$\pi_{d}F_{2}$& \multicolumn{1}{c}{}
$0$ & $0$ & $\ZZ \times \ZZ$ &  $\ZZ$ &  $0$ &  $\ZZ$ & $\ZZ_{6}$ & $0$ & $\ZZ_{12}$ & $\ZZ_{3}$ & $\ZZ_{30}$ &    \\
\end{tabular}
\caption{A list of the homotopy groups  $\pi_{d}F_{2}$ for dimensions $d \leq 10$. }
\end{table}

\begin{table}
\centering
\begin{tabular}{c|ccccccccc|cccccccc}
$d  $& \multicolumn{1}{c}{}
$0$ & $1$   &  $2$ &  $3$ &  $4$ &  $5$ & $6$ & $7$ &            \\
\hline
$\pi_{d}CP^{2}$& \multicolumn{1}{c}{}
$0$ & $0$ & $\ZZ$ &  $0$ &  $0$ &  $\ZZ$ & $\ZZ_{2}$ & $\ZZ_{24}$ &    \\
\end{tabular}
\caption{A list of the homotopy groups  $\pi_{d}CP^{2}$ for dimensions $d \leq 7$. }
\end{table}

${\it (i)}$ It is known \cite{KondoTaira} that nontriviality of $\pi_{2}F_{2}=\ZZ \times \ZZ$ accounts for the presence
of two different monopoles in the theory (cf. $\pi_{2}CP^{2}=\ZZ$ in Table II which
means that we deal with a monopole of one type). The second homotopy group is nontrivial due to the fact that the simply connected flag space $F_{2}$ is a compact symplectic manifold.

${\it (ii)}$ The integers in the RHS of $\pi_{3}F_{2}=\ZZ$
have the meaning of $SU(3)$ instanton topological charges because
$\pi_{3}F_{2}=\pi_{3}SU(3)=\pi_{3}SU(2)$.

${\it (iii)}$ The integers in the RHS of $\pi_{5}F_{2}=\ZZ$ describe some textures and
the corresponding phases. The nature of these textures is difficult to guess now.

${\it (iv)}$
The most interesting answer $\pi_{6}F_{2}=\ZZ_{6}$ means that there are only six phase states with the labels $\{0,1, \dots 5 \}$. They are usually
ordered as  three quark doublets. We can topologically distinct the quark states because of the gauge invariant coupling of the fermions to the gauge potential. This takes place on the scales where we can consider the six-dimensional base space as the sphere $\Sph^{6}$. Note that some additional parameters of the $(3+1)$d gauge theory can add dimensions in order to have finally 6 dimensions of the base space \cite{Volovik}. We encounter these phenomena in some topologically ordered phases of condensed matter \cite{TeoKane2}. In our case, the best natural choice
for the interpretation of the base space $\Sph^{6}$ corresponds to the standard six-dimensional space-momentum phase space. We are free also to interpret the six-dimensional compact space $\Sph^{6}$ as a complement to the $(3+1)$-dimensional space-time, but the previous suggestion is much better.

{\it Gauge fields on the flag space}. - Let us describe the flag space $F_{2}$ in detail to explain in particular at the end, why we addressed to homotopy theory approach. It is a compact K\"ahler manifold which is a homogeneous nonsymmetric space of dimension ${\rm dim} {F_{2}}=6$.
Since the flag manifold ${F_{2}}$ is the K\"ahler one, it possesses the complex local coordinates
$w_{\alpha}$, $\alpha =1,2,3$, the Hermitian Riemanian metric, $ds^2=g_{\alpha\bar \beta}dw^{\alpha}d\bar w^{\beta}$,
and the closed two-form (field strength)
$
%\begin{equation}
\Omega_K=i g_{\alpha\bar \beta}dw^{\alpha} \wedge d\bar w^{\beta}
%\label{K2f}
%\end{equation}
$
, i.e., $d\Omega_K=0$.  Here, $d = \partial + \bar \partial
 = dw_\alpha \frac{\partial}{\partial w_\alpha}
 + d\bar w_\beta \frac{\partial}{\partial \bar w_\beta}$ denotes the exterior derivative,
 while the operators $\partial$ and $\bar \partial$ are called the Dolbeault operators.

According to the Poincar\'e lemma, any closed form
$\Omega_K$ is {\it locally} exact, i.e., $\Omega_K= d\omega$, where $\omega$ is the gauge potential.
The condition $d\Omega_K=0$ is equivalent to $g_{\alpha\bar\beta} = \frac{\partial}{\partial w^{\alpha}}
\frac{\partial}{\partial \bar w^{\beta}} K ,$ where  $K=K(w,\bar w)$ is the K\"ahler potential:
\begin{eqnarray}
  K(w,\bar w)
  & =& \ln [(\Delta_1(w,\bar w))^m(\Delta_2(w,\bar w))^n] \, ,
\label{Kw}
\end{eqnarray}
\begin{eqnarray}
 \Delta_1(w,\bar w) & =& 1+|w_1|^2+|w_2|^2 \, ,
\end{eqnarray}
\begin{eqnarray}
\Delta_2(w,\bar w) = 1+|w_3|^2+|w_2-w_1 w_3|^2 \, .
\label{Delta}
\end{eqnarray}
By means of three complex variables $w_{\alpha}$, the flag space $F_{2}$ is realized as a set
of triangular matrices of the form
\begin{equation}
\left ( \begin{array}{ccc} 1 & w_1 & w_2\\0 & 1 & w_3\\0 & 0 & 1
\end{array}\right )^{t}
\in F_2 = SU(3)/U(1)^2   \, .
\label{eq:Matr}
\end{equation}

The K\"ahler one-form and the two-form are
$
%\begin{equation}
\omega=\frac{i}{2}(\partial - \bar\partial)K, \Omega_K  = i \partial \bar \partial K
%\label{OmegaK}
%\end{equation}
$.
The explicit forms of the gauge potential $\omega$ and the field strength $\Omega_K$ are given by
\begin{eqnarray}
\omega &=& im \frac{w_1d\bar w_1+w_2d\bar w_2}{\Delta_1(w, \bar w)} + in \frac{w_3d\bar w_3}
{\Delta_2(w, \bar w)}  \nonumber \\
&& + in \frac{w_2-w_1 w_3)(d\bar w_2-\bar w_1d\bar w_3-\bar
w_3d\bar w_1}
{\Delta_2(w, \bar w)} \, ,
\end{eqnarray}
\begin{eqnarray}
  \Omega_K &=&   d\omega
  = im (\Delta_1)^{-2}[(1+|w_1|^2)dw_2 \wedge d\bar w_2 \nonumber \\&&
  - \bar w_2 w_1 dw_2 \wedge d\bar w_1
  \nonumber\\&&
  - w_2 \bar w_1 dw_1 \wedge d\bar w_2
  + (1+|w_2|^2) dw_1 \wedge d\bar w_1]
  \nonumber\\&&
  + in (\Delta_2)^{-2}[\Delta_1 dw_3 \wedge d\bar w_3 \nonumber \\&&
  - (w_1+\bar w_3 w_2) dw_3 \wedge (d\bar w_2 - \bar w_3 d\bar w_1)
    \nonumber\\&&
  - (\bar w_1+w_3 \bar w_2)(dw_2-w_3 dw_1) \wedge d\bar w_3
  \nonumber\\&&
  + (1+|w_3|^2)(dw_2-w_3 dw_1)\nonumber \\&&
  (d\bar w_2 - \bar w_3 d\bar w_1)] \,.
  \label{K2fb}
\end{eqnarray}
Calculation of the Poincar\'e polynomial $P_{F_{2}}(t)=\sum\limits_{i=0}^{6}b_{i}t^{i}$ of $F_{2}$ (see \cite{Borel,MPS})
with the Betti numbers $b_{i}$ yields $P_{F_{2}}(t)=1+2t^{2}+2t^{4}+t^{6}$, i.e., $b_{0}=b_{6}=1, b_{2}=b_{4}=2$. We see
that the cohomology class is not zero because all even Betti numbers are nonzero.

The K\"ahler potential for $CP^2$ is given by
%$
\begin{equation}
  K(w,\bar w) = \ln [(\Delta_1)^m],
\end{equation}
%$
which is obtained as a special case of $F_2$ by specifying the coordinate $w_3=0$ and the parameter
$n=0$ in Eq.~(\ref{Kw}). Hence, we have
%$
\begin{eqnarray}
  \omega = im \frac{w_1d\bar w_1+w_2d\bar w_2}{\Delta_1(w, \bar w)}
\end{eqnarray}
%$
up to the total derivative and
\begin{eqnarray}
  \Omega_K &=&   d\omega
  = im (\Delta_1)^{-2}[(1+|w_1|^2)dw_2 \wedge d\bar w_2 \nonumber \\&&
  - \bar w_2 w_1 dw_2 \wedge d\bar w_1
  \nonumber\\&&
  - w_2 \bar w_1 dw_1 \wedge d\bar w_2 \nonumber \\&&
  + (1+|w_2|^2) dw_1 \wedge d\bar w_1] .
  \label{CP2K2f}
\end{eqnarray}
\par

This should be compared to the case $F_1=CP^1$, where
\begin{equation}
 e^{ w s_{+}}  =
\left(\begin{array}{cc}1 & w \\0 & 1\end{array}\right) \in F_1 = CP^1=
SU(2)/U(1) \cong {\Sph}^{2}
\label{s2}
\end{equation}
and
$
s_{+} =
\left(\begin{array}{cc}0 & 1 \\0 & 0\end{array}\right)
$. Here, the complex variable $w$ is the $CP^1$ variable written as
$
 w = e^{i\phi} \cot \frac{\theta}{2}
$
in terms of the polar coordinates of the unit vector ${\bf n}$ in ${\Sph}^{2}$.
The results for $SU(2)$ are well-known:
$
%\begin{eqnarray}
K(w,\bar w) = m \ln [(1+|w|^2)], \omega = i m \frac{wd\bar w}{1+|w|^2}
%\end{eqnarray}
$,
and
%\begin{eqnarray}
$
\Omega_K = i g_{w\bar w} dw \wedge d\bar w
 = i m  \frac{dw \wedge d\bar w}{(1+|w|^2)^{2}}
%\end{eqnarray}
$.

It is seen from these equations that the degenerate orbit $CP^{2}$ is the four-dimensional feature inside the six-dimensional flag space $F_{2}$.
Therefore,
%the
two-form (\ref{K2fb})
is closed, but not {\it globally} exact.
One can say that $\Omega_K$ is an element of the second cohomology group of $F_{2}$.
This means that we cannot define the gauge connection
$\omega$ everywhere in $F_{2}$, because the one-form $\omega$ is not well defined on the
manifold. This is the reason why it is difficult to directly determine the Hopf-like topological invariants in the $F_{2}$ case.

{\it  Topological invariants} . -  Let us proceed with the analysis of topological invariants by calculating the homotopy groups of $SU(3)$.
The flag space $F_{2}=SU(3)/U(1)^{2}$ is the base space of the $U(1)^{2}$-fiber bundle $SU(3)\rightarrow F_{2}$.
We have the following exact sequences:
\begin{eqnarray}
0 \rightarrow \pi_{d}\left(U(1)^{2}\right) \rightarrow \pi_{d}SU(3) \overset\cong\rightarrow \pi_{d}F_{2} \nonumber
\end{eqnarray}
\begin{eqnarray}
\rightarrow
\pi_{d}\left(U(1)^{2}\right) \rightarrow 0, \qquad {\text {for}} \,\, d \geq 3 \, ,
\end{eqnarray}
\begin{eqnarray}
0 \rightarrow \pi_{2}SU(3) \rightarrow \pi_{2}F_{2} \rightarrow \pi_{1}\left(U(1)^{2}\right) \nonumber
\end{eqnarray}
\begin{eqnarray}
\rightarrow
\pi_{1}SU(3) \rightarrow \pi_{1}F_{2} \rightarrow 0  \, ,
\end{eqnarray}
where $\pi_{1}\left(U(1)^{2}\right)=\ZZ \times \ZZ$ and $\pi_{1}SU(3)=\pi_{2}SU(3)=0$.
Thus, we have
\begin{eqnarray}
\left\{
\begin{array}{l@{\quad}l}
\pi_{0}F_{2}=\pi_{1}F_{2}=0,   \\
\pi_{2}F_{2}=\ZZ \times \ZZ,  \\
\pi_{d}F_{2}=\pi_{d}SU(3), \qquad {\text {for}} \,\, d \geq 3 \, .
\end{array}
\right .
\label{eq:newF2}
\end{eqnarray}

We summarized the results in Table I. It presents the nontrivial homotopy groups of $F_{2}$, which are in accord with previous studies \cite{MT} (see also Ref. \cite{DGMS}).  For completeness and comparison, we also have shown a list of the homotopy groups $\pi_{d}CP^{2}$ for $d \leq 7$ in Table II.

Table I is proved by the results of Ref. \cite{PuetRigas} and references therein.
In particular, Ref. \cite{PuetRigas} presents two theorems that account for the $5^{\rm th}$ and $6^{\rm th}$ homotopy groups
of SU(3).
{\it Theorem 1}: $\pi_{2n-1}U(N)=\ZZ$ for $N \geq n$ and {\it Theorem 2}: $\pi_{2n}U(n)=\ZZ_{n!}$ for $n \geq 2$.
We will study now the $3^{\rm rd}$, $5^{\rm th}$, and $6^{\rm th}$ homotopy groups in more detail.

{1. \it The $3^{\rm rd}$ homotopy group of SU(3)}. The exact sequence of the fibration  \\
$SU(3) \rightarrow SU(3)/SU(2)\cong \Sph^5$ is
\begin{eqnarray}
\pi_{d+1}\Sph^5 \rightarrow \pi_{d}SU(2) \overset{i_{*}}\rightarrow \pi_{d}SU(3) \rightarrow \pi_{d}\Sph^5 \,.
\label{exactseq}
\end{eqnarray}
Let $d=3$ and, since $\pi_{4}\Sph^5=\pi_{3}\Sph^5=0$, let the inclusion $i$: $SU(2)\hookrightarrow SU(3)$
induce an isomorphism $i_{*}$: $\pi_{3}SU(2) \overset\cong\rightarrow \pi_{3}SU(3)$. A generator for $\pi_{3}SU(2)$
is given by
\begin{eqnarray}
g_{2}({\bf r})=r^{0}{\bf 1} + ir^{j}\sigma_{j},
\label{generator}
\end{eqnarray}
\begin{eqnarray}
{\text {for}}\,\,\, {\bf r}= (r^{0},r^{1},r^{2},r^{3}) \,\,\, {\text
{and}} \quad |{\bf r}|=1  \, , \nonumber \\&&
\end{eqnarray}
where ${\bf 1}$ is the identity matrix and $\sigma_{j}$ are the Pauli matrices. Thus, the generator
for $\pi_{3}SU(3)$ is
\begin{equation}
g_{3}({\bf r})=\left ( \begin{array}{ccc} 1 & 0 & 0\\0 & r^{0}+ir^{3} & ir^{1}+r^{2}\\0 & ir^{1}-r^{2} & r^{0}-ir^{3}
\end{array}\right ) \, .
\label{eq:gen}
\end{equation}

Given any continuous function $g$: $\Sph^3 \rightarrow SU(3)$, the topological invariant,
i.e., the winding degree $[g] \in \pi_{3}SU(3)=\ZZ$, is determined by the integral formula
\begin{equation}
[g]=\frac{1}{24 \pi^{2}}\int_{\Sph^3} {\rm Tr}\left[(gdg^{\dag})^{3}\right]= \nonumber
\end{equation}
\begin{equation}
\frac{1}{24 \pi^{2}}\int_{\Sph^3}d^{3}x \varepsilon_{\mu\nu\lambda}\ {\rm Tr}(g^{\dag}\partial_{\nu}gg^{\dag}\partial_{\mu}gg^{\dag}\partial_{\lambda}g) \, .
\label{eq:gen2}
\end{equation}

{2. \it The $5^{\rm th}$ homotopy group of SU(3)}. Using exact sequence (\ref{exactseq}), we have
\begin{eqnarray}
\pi_{5}SU(2) \rightarrow \pi_{5}SU(3) \rightarrow \pi_{5}\Sph^5 \nonumber
\end{eqnarray}
\begin{eqnarray}
\rightarrow \pi_{4}SU(2)
\rightarrow \pi_{4}SU(3) \, .
\label{exactseq2}
\end{eqnarray}
It is known that $\pi_{5}SU(2)=\pi_{5}\Sph^3 = \ZZ_{2}$ and $\pi_{5}SU(3)=\ZZ$ (from theorem 1), and thus
the first arrow must be the zero map $(\ZZ_{2}\overset{0}\rightarrow \ZZ)$. We also know that
$\pi_{4}SU(3)=0$, as the homotopy group $\pi_{4} SU(N)=0$ stabilizes after $N \geq 3$. Finally, we need
$\pi_{4}SU(2)=\pi_{4}\Sph^3 =\ZZ_{2}$. We now have
\begin{eqnarray}
0 \rightarrow \pi_{5}SU(3) \overset{\times 2}\rightarrow \pi_{5}\Sph^5 \rightarrow \ZZ_{2}
\rightarrow 0 \, .
\label{exactseq3}
\end{eqnarray}
This means that given $g$: $\pi_{5}\Sph^5 \rightarrow SU(3)$,
\begin{eqnarray}
g({\bf r})=\left ( \begin{array}{ccc} {|} & {|} & {|}  \\
{\bf u}_{1}({\bf r}) & {\bf u}_{2}({\bf r}) & {\bf u}_{3}({\bf r})  \\
{|} & {|} & {|}
\end{array}\right ) \, , \quad {\text {for}} \, {\bf r} \in \Sph^5 ,
\label{eq:gen3}
\end{eqnarray}
the vector ${\bf u}_{1}$: ${\Sph}^{5} \rightarrow {\Sph}^{5}$ has an even
winding degree, namely, the winding degree $[{\bf u}_{1}]=2 \times {\text {winding degree}} [g]$.

The exact sequence for the fibration $SU(N+1) \rightarrow SU(N+1)/SU(N)={\Sph}^{2N+1}$ shows that
$\pi_{5}SU(N)=\ZZ$ stabilizes after $N \geq 3$. Thus,  the winding degree of $g$ can be also deduced by the usual formula
\begin{eqnarray}
[g]=\frac{1}{480\pi^{3}i}\int\limits_{{\Sph}^{5}} {\rm {Tr}}\left[(gdg^{\dag})^{5} \right] \,.
\label{eq:gen4}
\end{eqnarray}
A particular generator of $\pi_{5}SU(3)$ can be found in Ref.~\cite{PuetRigas}.

{3. \it The $6^{\rm th}$ homotopy group of SU(3)}. Exact sequence (\ref{exactseq}) yields
\begin{eqnarray}
\pi_{7}{\Sph}^{5} \rightarrow \pi_{6}SU(2) \rightarrow \pi_{6}SU(3) \nonumber
\end{eqnarray}
\begin{eqnarray}
\overset{0}\rightarrow \pi_{6}{\Sph}^{5} \overset{\cong}\rightarrow \pi_{5}SU(2) \, ,
\label{eq:exactseq4}
\end{eqnarray}
where $\pi_{7}{\Sph}^{5}=\ZZ_{2}$ and $\pi_{6}SU(2)=\ZZ_{12}$. It turns out that
$\pi_{6}SU(3)=\ZZ_{3!}=\ZZ_{6}$. A generator for $\pi_{6}SU(3)$ can be found in
\cite{PuetRigas} in page 6.

{\it Conclusion}. - In conclusion, we focus on the nontrivial
homotopy groups for $d \leq 6$ $\pi_{2}F_{2}$, $\pi_{3}F_{2}$, $\pi_{5}F_{2}$, and $\pi_{6}F_{2}$ considered so far for the spheres $\Sph^2$ as the base space. The generalization $\Sph^2 \to T^n$,
where $T^n$ is the $n$-dimensional torus, is an interesting and more complicated extension even
in the $SU(2)$ case. The result of calculations \cite{Pont} of the mapping class groups in the last case with $T^3=S^1\times S^1\times S^1$ leads to the linear superposition of the topological invariants beginning from the first Chern class to the Hopf invariant (see also \cite{Prot}). Similar behavior
also takes place in the case of $T^{3}=\Sph^{2}\times\Sph^{1}$. The classification problems of the mappings $T^{n} \to F_{2}$ are still totally open.

Up to now, we did not pay any attention to the relation between the existence of strong interaction in the  system and homotopy group results. It is well known in the condensed matter community that nontrivial answers $\ZZ_{2}$ or $\ZZ$ for the topological invariant of non-interacting systems change drastically in particular to $\ZZ_{8}$ in the case of the interacting system \cite{Kit}. Considering from this point of view the result $\pi_{6}(F_{2})=\ZZ_{6}$, one can say that we deal here with the significant interaction as it takes place in our QCD system.

Thorough understanding of the role of the flag space $F_{2}$ in the $SU(3)$ gauge theory is related to the search for an analog of the Hopf number, i.e., the linking number of pullbacks on a space $\MM$ of two arbitrary "points" on a target space $\NN$ of the map $\MM \rightarrow \NN$. Such an analog can have the form of pre-images of the target points in the codimension two \cite{Ran}. This could take place if $\MM = {\Sph}^{3}$ and $\NN$ is the $2d$ complement of $CP^{2}$ with respect to the whole space $F_{2}$. This is an open question, which is difficult to answer without knowing the
details of the map. We leave the problem of describing the details of this map for future work.

We are grateful to V.I. Arnold, J.E. Avron, J. Bernatska, L.D. Faddeev, C.L. Kane, Y. Hatsugai, A.W.W. Ludwig, I.A. Taimanov, G. E. Volovik  for fruitful discussions.
The coauthors (A.P., J.T.) especially thank the organizers of the 28th Jerusalem Winter School in Theoretical Physics A. Stern and S.-C. Zhang for hospitality at the Israel Institute for Advanced Studies where a part of the work was done.
\\\\
\noindent
${}^*$alprot@appl.sci-nnov.ru


\begin{thebibliography}{100}
\bibitem{Fa} The problem in this form has been formulated by L. Faddeev in Ref. \cite{Fad} (see also  Ref. \cite{BolFad}).
\end{thebibliography}
\end{document}